\documentclass[prl,floatfix,showpacs,aps,reprint,letterpaper,nobalancelastpage,superscriptaddress]{revtex4-1}

\usepackage{graphicx}
\usepackage{amsmath}
\usepackage{verbatim}
\usepackage{color}
\usepackage[urlcolor=blue, hyperindex, colorlinks, bookmarks=true,linkcolor=black,citecolor=black]{hyperref}

\newcommand{\dg}{^\dagger}

\begin{document}
\title{Superconducting qubit in waveguide cavity with coherence time approaching 0.1ms}
\author{Chad Rigetti}
\affiliation{IBM T. J. Watson Research Center, Yorktown Heights, NY 10598, USA}
\author{Stefano Poletto}
\affiliation{IBM T. J. Watson Research Center, Yorktown Heights, NY 10598, USA}
\author{Jay M. Gambetta}
\affiliation{IBM T. J. Watson Research Center, Yorktown Heights, NY 10598, USA}
\author{B. L. T. Plourde}
\affiliation{Department of Physics, Syracuse University, Syracuse, New York 13244-1130, USA}
\author{Jerry M. Chow}
\affiliation{IBM T. J. Watson Research Center, Yorktown Heights, NY 10598, USA}
\author{A. D. C\'orcoles}
\affiliation{IBM T. J. Watson Research Center, Yorktown Heights, NY 10598, USA}
\author{John A. Smolin}
\affiliation{IBM T. J. Watson Research Center, Yorktown Heights, NY 10598, USA}
\author{Seth T. Merkel}
\affiliation{IBM T. J. Watson Research Center, Yorktown Heights, NY 10598, USA}
\author{J. R. Rozen}
\affiliation{IBM T. J. Watson Research Center, Yorktown Heights, NY 10598, USA}
\author{George A. Keefe}
\affiliation{IBM T. J. Watson Research Center, Yorktown Heights, NY 10598, USA}
\author{Mary B. Rothwell}
\affiliation{IBM T. J. Watson Research Center, Yorktown Heights, NY 10598, USA}
\author{Mark B. Ketchen}
\affiliation{IBM T. J. Watson Research Center, Yorktown Heights, NY 10598, USA}
\author{M. Steffen}
\affiliation{IBM T. J. Watson Research Center, Yorktown Heights, NY 10598, USA}

\begin{abstract}
We report a superconducting artificial atom with an observed quantum coherence time of $T_{2}^{\ast}$=95$\mu$s and energy relaxation time $T_{1}$=70$\mu$s. The system consists of a single Josephson junction transmon qubit embedded in an otherwise empty copper waveguide cavity whose lowest eigenmode is dispersively coupled to the qubit transition. We attribute the factor of four increase in the coherence quality factor relative to previous reports to device modifications aimed at reducing qubit dephasing from residual cavity photons. This simple device holds great promise as a robust and easily produced artificial quantum system whose intrinsic coherence properties are sufficient to allow tests of quantum error correction.
\end{abstract}
\maketitle

Superconducting quantum circuits are a leading candidate technology for large scale quantum computing. They have been used to show a violation of a Bell-type inequality \cite{BellMart}; implement a simple two-qubit gate favorable for scaling \cite{chowIBM2Q}; generate three-qubit entanglement \cite{dicarlo3Q}; perform a routine relevant to error correction \cite{ReedQEC}; and very recently to demonstrate a universal set of quantum gates with fidelities greater than 95\% \cite{ChowUni}. Most of these devices employ small angle-evaporated Josephson junctions as their critical non-linear circuit components. Devices designs appear to be consistent with the basic requirements for quantum error correction (QEC) and fault tolerance \cite{DDsurf}. However, the construction and operation of much larger systems capable of meaningful tests of such procedures will require individual qubits and junctions with a very high degree of coherence. Current estimates for threshold error rates -- and the cumulative nature of errors originating from control, measurement, and decoherence -- make likely the need for quantum lifetimes \textit{at least} 10$^{3}$ times longer than gate and measurement times \cite{crossDD}, corresponding to 20 to 200$\mu$s for typical systems.

To this end, improvements in qubit lifetimes have continued for the past decade, spurred largely by clever methods of decoupling noise and loss mechanisms from the qubit transition and thus realizing Hamiltonians more closely resembling their idealized versions. Recently, Paik, \textit{et al}.  made a breakthrough advance \cite{Paik3D} by embedding a transmon qubit \cite{shreier, kochtransmon} in a superconducting waveguide cavity. Dubbed \textit{three-dimensional circuit QED} (3D cQED), this system produced significantly enhanced qubit lifetimes of T$_{1}$=25--60$\mu$s and T$_{2}^{\ast}$=10--20$\mu$s, corresponding to quality factors for dissipation and decoherence of $Q_{1}\approx$1.8$\times $10$^{6}$ and $Q_{2}\approx$7$\times $10$^{5}$, respectively.

These results lead to two important questions. First, are similar coherence properties observable using other fabrication processes, facilities, and measurement setups? Second, what is the origin of the dephasing process suppressing T$_{2}^{\ast }$ well below the no-pure-dephasing limit of 2T$_{1}$? Is it intrinsic to the junctions or to this qubit architecture? The weight and urgency of these questions are increased by implications on scaling potential: if the results are reproducible and decoherence times can be extended close to the 2T$_{1}$ limit for observed T$_1$ times, this technology becomes a strong candidate for the immediate construction of prototype processors for testing QEC without significant need of longer coherence. It would also suggest that other designs employing small angle-evaporated junctions (e.g. traditional planar integrated superconducting circuits) could potentially attain similar coherence if present performance limits can be identified and overcome.

In this Letter we report a 3D cQED device that demonstrates the basic reproducibility of Paik,\textit{\ et al.} and, moreover, shows that decoherence times can be extended further by taking precautions against qubit dephasing induced by fluctuations of the cavity photon number during qubit operation. With such precautions our system was observed to have $T_{2}^{\ast }$=95$\mu$s ($Q_{2}\approx $2.5$\times$10$^{6}$) and $T_{1}=70\mu$s ($Q_{1}\approx $1.8$\times $10$^{6}$). This level of performance places our device already well within the regime where standard microwave techniques should allow quantum gate fidelities exceeding those required for fault tolerance.

In the 3D cQED framework, qubits are manufactured with standard lithographic processes while the cavities, simple macroscopic resonant enclosures, are made independently and with very different techniques, such as precision machining or laser etching. Individual qubits and cavities can be treated as discrete components; their properties, materials, and designs may be varied independently without special effort. The large mode volumes, structural simplicity, and absence of very large aspect ratio thin films make full-device electromagnetic simulation highly accessible. These properties in turn facilitate a high degree of practical control and engineerability of the qubit system and its electromagnetic environment.

In this work we exploit these and other properties of 3D cQED systems to engineer a device more robust against qubit dephasing due to the presence and fluctuations of residual photon population in the cavity. We do this by three parallel strategies. First, we select qubit and cavity parameters to reduce the expected qubit dephasing rate per residual photon in the fundamental cavity mode. Second, we engineer the device to limit the spectral proximity of and couplings to higher modes of the cavity. We employ a symmetric cavity shape such that the next nearest mode after the fundamental that couples to the qubit is at $\approx$24 GHz, or more than 20 GHz away from the qubit transition. This design minimizes the role of higher modes and makes the standard single-mode cavity approximation more robust. Third, we aim to suppress residual cavity population by following a simple rule: the thermal photon temperature of a resonant mode, be it linear or nonlinear, is bounded by the temperature of the dissipation source limiting its quality factor ($Q$). Typically, $Q$'s of cQED resonators are limited by the ohmic environment external to the resonator to which the device is coupled. But it is notoriously difficult to ensure that the modes of a feedline are thermalized to very low temperatures \cite{JerryNumSplt}. Rather than solving this problem directly, we instead make use of an ideal cold resistor -- the interior walls of a bulk oxygen-free high-conductivity (OFHC) copper cavity -- as the primary source of dissipation. In conjunction with the under-coupling of the cavity to the external environment, internal cavity dissipation is expected to thermalize the cavity population to the temperature of the bulk copper, which in turn is easily anchored to the lowest available temperature.

\begin{figure}[t!]
\centering
\includegraphics[width=0.48\textwidth]{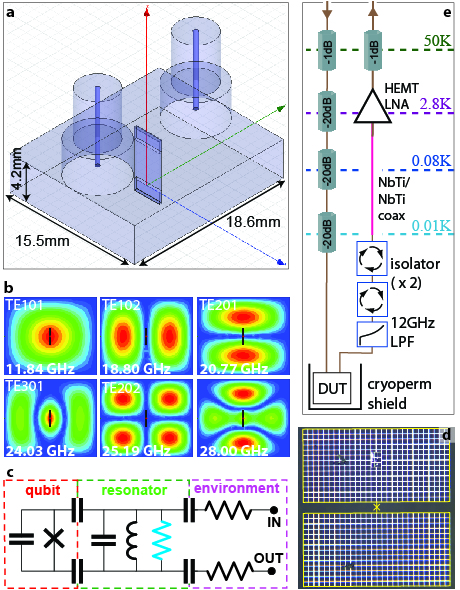}
\caption{\label{fig:1} (color online) Transmon qubit in three-dimensional copper waveguide cavity with long coherence. (a) Device model (HFSS) showing interior volume of the waveguide enclosure housing a sapphire chip and transmon qubit, with two symmetric coaxial connectors for coupling signals in/out. (b) Eigenmodes of the enclosure with sapphire chip (obtained with HFSS eigenmode solver) illustrating robustness of single-mode-cavity approximation. The qubit is positioned at a cavity symmetry point where the fundamental mode (TE101) is maximal. Device dimensions and symmetries imply that next mode interacting with qubit (TE301) is $>$ 20 GHz detuned from qubit. (c) Equivalent circuit diagram of device. The interior walls of copper waveguide cavity provide a readily thermalized cold resistor (light blue) that is expected to sink the residual cavity photon population to the lowest available temperature. (d) Optical image of transmon, consisting of two capacitor pads 350$x$ 700$\mu$m$^2$ each (outlined in yellow) separated by a $50\mu$m wire interrupted by a shadow evaporated Al/AlOx/Al Josephson junction (yellow overlay). Pads are formed of mesh with $5\mu$m wires and 20$\mu$mx20$\mu$m holes to suppress vortex trapping and motion. (c) Cryogenic microwave measurement setup. Experiments performed at base temperature of 8mK in a BlueFors cryogen-free dilution refrigerator.}
\end{figure}

The three-dimensional circuit QED sytem we report is described in the two-level, dispersive, and single-mode cavity approximations by the Hamiltonian \cite{Gambetta2006}
\begin{equation}
H/\hbar =\omega _{c}a^{\dag }a-\frac{\omega_{01}}{2}\sigma _{z} - \chi a^{\dag} a\sigma_z.
\end{equation}
The term proportional to $a^{\dag} a\sigma_z$ can be interpreted as a cavity photon number dependent shift of the qubit transition frequency (ac-Stark shift), or as a qubit state dependent shift of the cavity frequency (cavity pull). For transmons the cavity pull $\chi =-g^{2}E_{C}/(\Delta^2-\Delta E_{C})$ where $g$ is the bare coupling strength, $\Delta =\omega _{01}-\omega _{c}$ is the cavity--qubit detuning and $E_{C}=e^{2}/2C_{\Sigma}$ is the transmon charging energy and $C_{\Sigma}$ is the total qubit capacitance \cite{kochtransmon}. In the strong dispersive coupling regime the cavity pull can be larger than the intrinsic linewidth of the qubit transition. In such systems, fluctuations of the cavity photon number scramble the qubit frequency and place a limit on coherence. This occurs through the same mechanism that allows the cavity photons to induce a projective measurement of the qubit state \cite{Gambetta2006}. It can result from both thermal and coherent cavity photon populations.

We represent thermal driving of the resonator by the master equation \cite{Gardiner2004a}
\begin{equation}
\begin{split}
\dot\rho  = -\frac{i}{\hbar}[H,\rho] +\kappa_{j}(n_{\mathrm{th}_{j}}+1) \mathcal{D}[a]\rho
+\kappa_{j}(n_{\mathrm{th}_{j}}) \mathcal{D}[a\dg]\rho,
\end{split}
\label{eq_master_eq}
\end{equation}
where $\mathcal{D}[\hat L]\rho = \left(2 \hat L \rho \hat L^\dag
-\hat L^\dag \hat L \rho - \rho \hat  L^\dag \hat L\right)/2$ is the Lindblad super-operator for dissipation, $\kappa_{j}$ is the cavity relaxation rate through source $j$, and
$n_{\mathrm{th}_{j}} = 1/(e^{\hbar\omega_{01}/kT_j}-1)$ is the thermal photon number for source $j$ at temperature $T_j$.
Following a similar procedure to Refs. \cite{Dykman1987,Clerk2007} the thermal-induced dephasing rate at times long compared to $1/\kappa_\mathrm{tot}$, where $\kappa_\mathrm{tot}$ is the cumulative dephasing rate, is,
\begin{equation}
\Gamma_\mathrm{th} =\frac{\kappa_\mathrm{tot}}{2} \mathrm{Re} \left[\sqrt{\left(1+\frac{2i \chi}{\kappa_\mathrm{tot}}\right)^2 +\left(\frac{8i\chi\sum_j\kappa_jn_{\mathrm{th}_{j}} }{\kappa_\mathrm{tot}^2}\right)}-1\right].
\end{equation} For large $\kappa_\mathrm{tot}/\chi$ \cite{Bertet2005},
\begin{equation}
\Gamma_\mathrm{th} = \frac{4\chi^2\sum_j\kappa_jn_{\mathrm{th}_{j}}}{\kappa_\mathrm{tot}^2}\left(\sum_j\kappa_jn_{\mathrm{th}_{j}}+1\right)
\end{equation} while it saturates to  $\Gamma_\mathrm{th} = \sum_j\kappa_jn_{\mathrm{th}_{j}}$ for large $\chi/\kappa_\mathrm{tot}$.

These suggest different possible strategies to mitigate cavity photon induced dephasing: suppress fluctuations of the photon number by using very high $Q$ cavities, effectively pushing photon shot noise to lower and lower frequencies; or, suppress the photon number with a cold dissipation source internal to the cavity. In the first strategy, one must ensure that the modes of the feedlines coupled to the cavity are cold at all relevant frequencies, as the cavity modes will thermalize to the same temperature. In the second, the internal cold dissipation is expected to thermalize all cavity modes. In both cases, one pays a price in effective signal to noise of the measurement, but for different reasons. The cold dissipation leads to a loss of information-carrying photons within the cavity before they can be measured; the high $Q$ approach requires longer and longer measurement integration and repetition times.

Here we follow the second strategy by employing the interior surfaces of an enclosure machined from bulk OFHC copper as an ideal cold resistor that appears as parallel damping of the effective cavity resonant circuit and limits its $Q$. The cavity, accordingly, is under-coupled to the input and output transmission lines.

\begin{figure}[t!]
\centering
\includegraphics[width=0.48\textwidth]{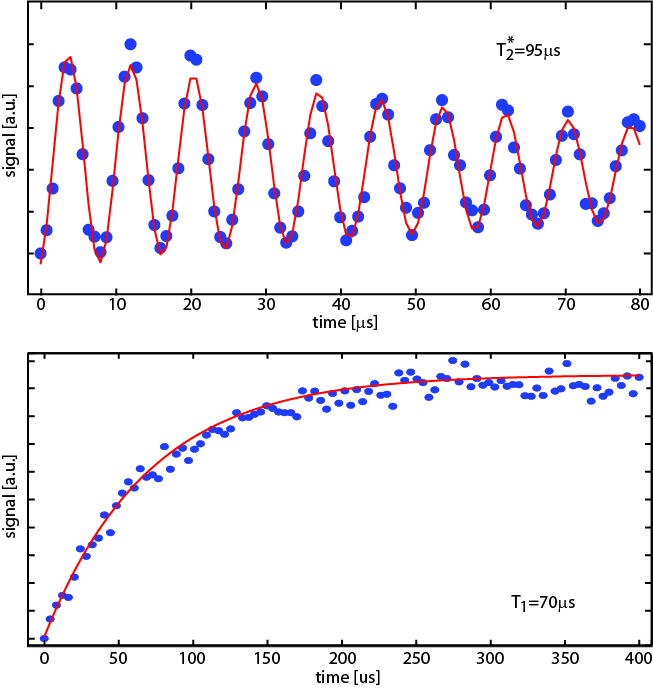}
\caption{\label{fig:1} (color online) Quantum state lifetimes. (a) Ramsey fringe experiment to determine coherence time. (b) Energy relaxation time.}
\end{figure}

Our device is shown in figure 1. The transmon qubit junction has characteristic energy $E_{J}/h$=10.3946 GHz. The qubit capacitance is determined from measurement to be $C_{\Sigma }$=91fF, implying $E_{J}/E_{C}$=49, placing the qubit in the transmon regime \cite{Gambetta2006, shreier, kochtransmon}. Qubits are produced in a 3" wafer process on c-plane 330$\mu$m thick sapphire prior to dicing into 3.2mm$\times$6.7mm chips. The chip is enclosed in a cavity machined from bulk OFHC copper. The cavity is formed by two halves and has in the assembled state an interior volume 18.6$\times $15.5$\times $4.2mm$^{3}$ plus symmetric cylindrical perturbations ($d$=7.7mm; $h$=4mm) of the ceiling to accommodate commercial bulkhead SMA connectors through which signals are coupled in and out. Measurement is performed with standard circuit QED dispersive measurement techniques \cite{BlaiscQED} using the lowest resonant mode of the enclosure (TE101) which is found experimentally at 12.1 GHz. The cavity is closed with brass screws, wrapped with Eccosorb foam and aluminized mylar to protect against stray radiation \cite{tony}, placed inside a cryoperm magnetic shield, and thermalized to the mixing chamber stage of a dilution refrigerator.

Micrwoave signals are supplied from room temperature electronics to the cavity through standard attenuated wideband coaxial lines. Measurement signals exiting the cavity pass through (respectively) a 12 GHz low pass filter (K\&L) thermalized to 10 mK with a copper wire wrap; two double-junction and magnetically shielded isolators (Pamtech) thermalized to 10 mK via large copper plates with multiple high-pressure contact points; a NbTi/NbTi superconducting cable from 10mK to the 2.8 K stage and thermalized at each end and at its midpoint (80mK) with wire wrap. At 2.8 K the signal is amplified by a low-noise wideband HEMT amplifier (Caltech) operating from 6-18 GHz with a noise temperature of 10-15 K. The signal is amplified again at room temperature before being mixed down to 10MHz and digitized. In-phase and quadrature components of the signal are summed to produce a measurement of the qubit energy eigenstate.

We performed standard measurements to characterize qubit device properties and performance. We find the following parameters: $\omega _{01}/2\pi\approx$4.2 GHz; $\omega _{c}/2\pi\approx$12.1 GHz; $g/2\pi=$153 MHz; $\chi$=390 kHz; and a cavity with an internally limited $Q$=10,400 \cite{little note} implying a Purcell limit on qubit energy relaxation time of about 400$\mu$s, well above the measured T$_{1}$. Excited state lifetime and Ramsey fringe experiments yield $T_{1}$ = 70$\mu s$ and $T_{2}^{\ast}$=95$\mu$s (Figure 2). Our data are consistent with our hypothesis that a 3D circuit QED system whose cavity has a lower internal quality factor can facilitate significantly improved qubit coherence.

To test that cold dissipation and under-coupling to the feedline play a significant role, we measured a sister device (same cavity design; qubit fabricated on same wafer) with an over-coupled feedline. Consistent with expectations and Eq. (3) we observed significantly enhanced dephasing, with $T_{1}$=45$\mu$s and $T_{2}^{\ast}$=18$\mu$s, corresponding to a ~10x increase in pure dephasing rate. Further tests are necessary to verify the validity of Eq. (3).

What are the trade-offs with this approach? The experimenter pays a price in convenience: the increased internal dissipation implies that for every photon exiting the cavity to be amplified and measured three are dissipated in the normal metal walls. This places greater demands on the performance of the amplification chain to achieve a particular signal-to-noise ratio of the measurement.

An ideal setup would employ transmission lines whose modes are already thermalized to the lowest available temperature. This is indeed the objective but can be challenging to realize in practice due to basic materials properties at mK temperatures and the sensitivity of the system to even small fractions of a photon. In one experiment known to the authors, great care was taken to reduce this temperature as much as possible and a bound of about 55mK was achieved \cite{JerryNumSplt}. Reliably achieving mode temperatures even that low is a major challenge and can be difficult to reproduce from one experimental setup to another. For this reason we have instead taken the multi-pronged approach described in this Letter.

An auxiliary benefit of the bulk copper cavity is the reliability of the thermal link between the qubit substrate and the coldest temperature stage of the fridge. An interesting avenue for future work entails controlled design of the cold dissipation in the cavity such that higher overall cavity $Q$'s are obtained while otherwise maintaining the properties of the device we've described here. This could be done, for example, with bulk copper cavities whose interior walls are partially coated with a thin layer of aluminum; we have begun work towards this.

We have constructed a 3D qubit system based on a single-junction transmon in a copper waveguide cavity with lifetimes $T_{2}^{\ast}$=95$\mu$s and $T_{1}$=70$\mu$s. Our results provide evidence that highly coherent superconducting qubits based on small shadow-evaporated Josephson junctions are reproducible with different fabrication processes, facilities. By pursuing three parallel approaches to improving coherence limits due to cavity photon induced dephasing we attained a factor of four improvement in coherence quality factor Q$_{2}$=2.5$\times $10$^{6}$ relative to previous reports. Our device falls well within the range of performance required for large scale tests of error correction and fault tolerant quantum computing procedures. We believe this performance, along with the simplicity and discrete nature of the qubits and cavities, makes this technology a strong candidate for the immediate construction of prototype quantum processors with 10-1000 qubits.

We acknowledge support from IARPA under Contract No. W911NF-10-1-0324. We acknowledge discussions and contributions from  Jack Rohrs, Joel Strand, Matthew Ware and Michael DeFeo. All statements of fact, opinion or conclusions contained herein are those of the authors and should not be construed as representing the official views or policies of the U.S. Government.


\begin{thebibliography}{99}
\bibitem{little note} Identical aluminum cavities with an internal Q$>$4M were measured to have Q=30,00-40,000 with the same connector configuration.
\end{thebibliography}
\end{document}